\documentclass[aps,prl,twocolumn,nofootinbib,showpacs,longbibliography]{revtex4-2}
\usepackage{bm,amsmath,amssymb,graphicx}
\usepackage[colorlinks,linkcolor=blue,urlcolor=blue]{hyperref}
\begin{document}

\title{Conserved Currents for the Gauge-Field Theory with Lorentz Symmetry Group}

\author{Hans Christian \"Ottinger}
\email[]{hco@mat.ethz.ch}
\homepage[]{www.polyphys.mat.ethz.ch}
\affiliation{ETH Z\"urich, Department of Materials, CH-8093 Z\"urich, Switzerland}

\date{\today}

\begin{abstract}
For the Yang-Mills-type gauge-field theory with Lorentz symmetry group, we propose and verify an explicit expression for the conserved currents in terms of the energy-momentum tensor. A crucial ingredient is the assumption that the gauge symmetry arises from the decomposition of a metric in terms of tetrad variables. The currents exist under the weak condition that the energy-momentum tensor and the Ricci tensor commute. We show how the conserved currents can be used to obtain a composite theory of gravity and discuss the static isotropic field around a point mass at rest.
\end{abstract}

% \pacs{04.50.Kd}
% 04.50.Kd   Modified theories of gravity

\maketitle

Soon after the original work of Yang and Mills on gauge-field theories \cite{YangMills54}, Utiyama \cite{Utiyama56} was interested in the theory with Lorentz symmetry group because there exist remarkable connections to Einstein's theory of gravity. The ideas of Utiyama's pioneering work, which may be considered as the origin of what is now known as gauge gravitation theory \cite{CapozzielloDeLau11,IvanenkoSar83}, have been criticized as ``unnatural'' by Yang (see footnote~5 of \cite{Yang74}). Yang's own, allegedly more natural proposal \cite{Yang74} has itself been criticized massively in Chapter~19 of \cite{BlagojevicHehl}. Further clarification of the relevance of the Yang-Mills theory with Lorentz symmetry group to the theory of gravity is still needed, most importantly to benefit from the experience with quantizing Yang-Mills theories in the context of gravity.

The present work addresses the unsolved fundamental problem of identifying the conserved currents associated with the Lorentz symmetry group, ideally in terms of the energy-momentum tensor. The proposed currents imply the proper coupling of the gravitational field to matter in a composite theory of gravity \cite{hco231,hco240}.

Yang-Mills theories are characterized by Lorentz covariant field equations with additional continuous local symmetries, formulated in a background Minkowski space-time. We denote the Minkowski metric with signature $(-,+,+,+)$ by $\eta_{\kappa\lambda} = \eta^{\kappa\lambda}$. Unless stated otherwise, the Minkowski metric is used for raising or lowering space-time indices throughout this paper.

The basic vector fields $A_{a \nu}$ of a Yang-Mills theory are labeled by two indices. In addition to the space-time index $\nu$, there is an index $a$ that labels a set of base vectors of the Lie algebra associated with an underlying continuous symmetry group. Yang-Mills theories are usually considered for the compact special unitary group ${\rm SU}(N)$ of unitary complex $N \times N$ matrices with determinant $1$ and the corresponding Lie algebra ${\rm su}(N)$, which is a vector space of dimension $N_{\rm p}=N^2-1$. In other words, the index $a$ takes values from $1$ to $N_{\rm p}$. Note that $N_{\rm p}$ may also be regarded as the number of continuous parameters required to characterize the elements of the underlying Lie group.

The elegance of the Yang-Mills equations for the vector fields $A_{a \nu}$ stems from the fact that they are covariant both under global Lorentz transformations (associated with the label $\nu$ of the vector fields) and under infinitesimal local symmetry transformations, or gauge transformations, from the underlying group (associated with the label $a$ of the gauge fields). We here assume that the local symmetry transformations are given by the Lorentz group ${\rm SO}(1,3)$ with Lie algebra ${\rm so}(1,3)$. This group consists of all the real $4 \times 4$ matrices that leave the Minkowski metric invariant, that is, of the matrices representing rotations in space or Lorentz boosts mixing space and time. The total number of parameters for the Lorentz group is $N_{\rm p}=6$, three for characterizing rotations and three for the boosts. The following choice of the six base vectors of the Lie algebra is quite natural: three generators of boosts in the three spatial directions ($a=1,2,3$) and three generators of rotations around the three coordinate axes ($a=4,5,6$). More illuminating than the labels $a$ are the corresponding pairs of space-time indices $(\kappa,\lambda)$ given in Table~\ref{tabindexmatch}, which refer to the two-dimensional subspaces in which the corresponding generators act nontrivially. Note that the boosts lead from the compact group of rotations ${\rm SO}(3)$ to the non-compact Lorentz group ${\rm SO}(1,3)$.

\begin{table}
\begin{tabular}{c|c c c c c c}
	% \hline
    $a$ \, & \, $1$ & $2$ & $3$ & $4$ & $5$ & $6$ \\
	\hline
	$(\kappa,\lambda)$ \,
    & \, $(0,1)$ & $(0,2)$ & $(0,3)$ & $(2,3)$ & $(3,1)$ & $(1,2)$ \\
	% \hline
\end{tabular}
\caption{Correspondence between the label $a$ for the base vectors of the six-dimensional Lie algebra ${\rm so}(1,3)$ of the Lorentz group and ordered pairs $(\kappa,\lambda)$ of space-time indices.}
\label{tabindexmatch}
\end{table}

Given any six components $\Lambda_a$, the quantities $\Lambda_{(\kappa,\lambda)}$ are directly defined for the index pairs given in Table~\ref{tabindexmatch}. It is natural to define $\Lambda_{(\kappa,\lambda)}$ for all pairs of space-time indices by antisymmetric continuation. The antisymmetric matrix  $\Lambda_{(\kappa,\lambda)}$ with six degrees of freedom contains no more and no less information than the original $\Lambda_a$.

Following standard procedures for Yang-Mills theories (see, e.g., Sect.~15.2 of \cite{PeskinSchroeder}, Chap.~15 of \cite{WeinbergQFT2}, or \cite{hco229}), we can introduce a field tensor $ F_{a \mu\nu}$ in terms of the gauge vector fields,
\begin{equation}\label{Fdefinition}
   F_{a \mu\nu} = \frac{\partial A_{a \nu}}{\partial x^\mu}
   - \frac{\partial A_{a \mu}}{\partial x^\nu}
   + f^{bc}_a A_{b \mu} A_{c \nu} ,
\end{equation}
where $f^{bc}_a$ stands for the structure constants of the Lie algebra, here for the Lorentz group, and the coupling constant is taken as unity. A Lie algebra label, say $a$, can be raised or lowered by raising or lowering the indices of the pairs associated with $a$ according to Table~\ref{tabindexmatch} by means of the Minkowski metric. The structure constants for ${\rm so}(1,3)$ can then be specified as follows: $f^{abc}$ is $1$ ($-1$) if $(a,b,c)$ is an even (odd) permutation of $(4,5,6)$, $(1,3,5)$, $(1,6,2)$ or $(2,4,3)$, and $0$ otherwise. The field equations of Yang-Mills theory can now be written as
\begin{equation}\label{YMfieldeqs}
   \frac{\partial F_{a \mu\nu}}{\partial x_\mu} + f_a^{bc} \, A_b^\mu F_{c \mu\nu} = - J_{a \nu} ,
\end{equation}
where the vector fields $J_{a \nu}$ are referred to as external sources or conserved currents. Indeed, by acting with the derivative $\partial/\partial x_\nu$ on Eq.~(\ref{YMfieldeqs}) and using the properties of structure constants, one can derive the conservation laws
\begin{equation}\label{Jaconservation}
   \frac{\partial J_{a \nu}}{\partial x_\nu} + f_a^{bc} \, A_b^\nu J_{c \nu} = 0.
\end{equation}
Our goal is to identify the conserved currents $J_{a \nu}$ associated with the Lorentz symmetry group.

For the construction of the currents it is crucial to understand how the Lorentz gauge symmetry arises. We consider the following decomposition of a symmetric $4 \times 4$ matrix $g_{\mu\nu}$,
\begin{equation}\label{localinertialdecomp}
   g_{\mu\nu} = \eta_{\kappa\lambda} \, {b^\kappa}_\mu {b^\lambda}_\nu
   = {b^\kappa}_\mu \, b_{\kappa\nu} .
\end{equation}
The gauge degrees of freedom arise from the fact that we consider only the matrix $g_{\mu\nu}$ as physical, not the individual factors ${b^\kappa}_\mu$ in the decomposition (\ref{localinertialdecomp}). The invariance properties of the Minkowski metric $\eta_{\kappa\lambda}$ imply that ${b^\kappa}_\mu$ can be multiplied from the left with an arbitrary Lorentz transformation matrix without changing $g_{\mu\nu}$.

Equation (\ref{localinertialdecomp}) may be regarded as a local transformation of the Minkowski metric to the background coordinate system. Therefore, the symmetric matrix $g_{\mu\nu}$ has all the properties required for a metric in general relativity, and the quantities ${b^\kappa}_\mu$ can be regarded as tetrad or \emph{vierbein} variables. More intuitively, the tetrad variables establish a relation between the underlying Minkowski coordinate system and freely falling local coordinate systems. However, the metric cannot be used in the same fully geometric interpretation as in general relativity because the Yang-Mills theory is defined in a fixed background Minkowski space-time. The deeper physical significance of the metric $g_{\mu\nu}$ is that it provides an anisotropic relationship between the four-vectors of velocity and momentum.

A two-fold role of the Lorentz group can be recognized nicely in the context of the tetrad variables ${b^\kappa}_\mu$. Lorentz transformations can independently act on the space-time indices $\kappa$ and $\mu$. On the one hand, when a Lorentz transformation acts on the index $\kappa$, it represents a local symmetry transformation leaving the metric (\ref{localinertialdecomp}) invariant. On the other hand, when a Lorentz transformation acts on the index $\mu$, it represents a global change of the coordinate system in the underlying Minkowski space. The index $\mu$ makes ${b^\kappa}_\mu$ a set of Lorentz four-vector fields, the index $\kappa$ introduces internal degrees of freedom associated with the Lorentz group as symmetry group.

Einstein's principle of equivalence of gravitational and inertial mass suggests a geometrization of the gravitational interaction, but not necessarily in terms of pseudo-Riemannian space-time geometry \cite{Jimenezetal19}. The metric $g_{\mu\nu}$ may merely be used to introduce an anisotropic velocity-momentum relation. In other words, the metric simply provides a relation between the local tangent and cotangent spaces. In the spirit of the equivalence principle, the tetrad variables ${b^\kappa}_\mu$ can be interpreted as a mapping between the anisotropic velocity-momentum relation in the background Minkowski space and an isotropic velocity-momentum relation in a freely falling local coordinate system. In that sense, the anisotropy effect of gravity is eliminated in the freely falling system. A gauge theory arises from Lorentz transformations in the local freely falling system, and the metric becomes a tensor under global Lorentz transformations in the background Minkowski space. Although the full space-time geometry is eliminated, the equation of motion for a particle in a composite gauge field may still be of the geodesic type (see Section~VI of \cite{hco231} and Sections IV A and G of \cite{hco240}).

We now express the vector fields of the Yang-Mills theory in terms of the tetrad variables of the decomposition (\ref{localinertialdecomp}) such that the former variables inherit the proper gauge transformation behavior from the latter ones. We thus arrive at a so-called composite Yang-Mills theory. A general Hamiltonian approach for composite theories has been developed in \cite{hco235,hco237}. The basic idea of composite gravity \cite{hco231,hco240} is to express the components of the gauge vector fields $A_{(\kappa\lambda) \rho}$ in terms of the tetrad fields ${b^\kappa}_\mu$ and their derivatives as follows,
\begin{eqnarray}
   A_{(\kappa\lambda) \rho} &=&
   \frac{1}{2} \, \mbox{$\bar{b}^\mu$}_{\kappa} \left( \frac{\partial g_{\nu\rho}}{\partial x^\mu}
   - \frac{\partial g_{\mu\rho}}{\partial x^\nu} \right) \mbox{$\bar{b}^\nu$}_{\lambda}
   \nonumber\\
   &+& \frac{1}{2} 
   \left(\frac{\partial b_{\kappa\mu}}{\partial x^\rho} \, \mbox{$\bar{b}^\mu$}_{\lambda}
   - \mbox{$\bar{b}^\mu$}_{\kappa} \, \frac{\partial b_{\lambda\mu}}{\partial x^\rho} 
   \right) , \qquad 
\label{compositionrule}
\end{eqnarray}
where $\mbox{$\bar{b}^\mu$}_{\kappa}$ is the inverse of the regular matrix ${b^\kappa}_\mu$. The Yang-Mills field equations (\ref{YMfieldeqs}) then become third-order differential equations for the tetrad fields, where the conserved currents $J_{a \nu}$ still remain to be found.

We now construct the currents $J_{a \nu}$ in terms of the energy-momentum tensor ${T_\mu}^\nu$, which satisfies the conservation laws
\begin{equation}\label{enmomconservation}
   \frac{\partial{T_\mu}^\nu}{\partial x^\nu} = \Gamma^\rho_{\mu\nu} {T_\rho}^\nu ,
\end{equation}
where $\Gamma^\rho_{\mu\nu}$ denotes the Christoffel symbols characterizing the affine metric connection. Compared to general relativity, a term $-\Gamma^\nu_{\nu\rho} {T_\mu}^\rho$ describing volume changes associated with the determinant of the metric is missing because we work in a background Minkowski space. As the energy-momentum tensor is gauge invariant, it is easier to construct the equivalent but gauge invariant current variables
\begin{equation}\label{Jtildedeftf}
   \tilde{J}^\rho_{\mu\nu} = {b^\kappa}_\mu {b^\lambda}_\nu \, J^\rho_{(\kappa\lambda)} ,
\end{equation}
obtained by transformation of the currents from the local freely falling system to the background Minkowski space. By means of a useful identity following from the values of structure constants associated with the Lorentz group,
\begin{equation}\label{supauxf}
   f_{(\kappa\lambda)}^{bc} B_b C_c =
   \eta^{\kappa'\lambda'} \Big[ B_{(\kappa'\lambda)} C_{(\kappa\lambda')}
   - C_{(\kappa'\lambda)} B_{(\kappa\lambda')} \Big] ,
\end{equation}
for arbitrary fields $B_b$ and $C_c$, the conservation law (\ref{Jaconservation}) for the currents $J_{a \nu}$ can be transformed into the equivalent conservation law
\begin{equation}\label{Jtildeconservation}
   \frac{\partial \tilde{J}_{\mu\nu}^\rho}{\partial x^\rho}
   - \Gamma^\sigma_{\rho\mu} \tilde{J}_{\sigma\nu}^\rho
   - \Gamma^\sigma_{\rho\nu} \tilde{J}_{\mu\sigma}^\rho = 0 .
\end{equation}

If we shift the energy-momentum tensor by an isotropic term,
\begin{equation}\label{Sdefinition}
   {S_\mu}^\nu = {T_\mu}^\nu - \frac{1}{2} {\delta_\mu}^\nu \, {T_\sigma}^\sigma ,
\end{equation}
the explicit expression for the transformed currents satisfying the conservation laws (\ref{Jtildeconservation}), which is the main result of the present work, can be written in the transparent form
\begin{eqnarray}
   \tilde{J}_{\mu\nu}^\rho &=& \frac{8\pi G}{c^4} \bigg(
   \frac{\partial {S_\nu}^\rho}{\partial x^\mu} + \Gamma^\rho_{\mu\sigma} {S_\nu}^\sigma
   - \Gamma^\sigma_{\sigma\mu} {S_\nu}^\rho \nonumber \\
   && \hspace{2em} - \, \frac{\partial {S_\mu}^\rho}{\partial x^\nu} 
   - \Gamma^\rho_{\nu\sigma} {S_\mu}^\sigma
   + \Gamma^\sigma_{\sigma\nu} {S_\mu}^\rho \bigg) , \qquad 
\label{Jtildedef}
\end{eqnarray}
where $G$ is Newton's constant and $c$ is the speed of light. By inserting this proposed explicit form of the currents into the left-hand side of Eq.(\ref{Jtildeconservation}), we obtain the following result after many cancellations,
\begin{eqnarray}
   \frac{\partial \tilde{J}_{\mu\nu}^\rho}{\partial x^\rho}
   - \Gamma^\sigma_{\rho\mu} \tilde{J}_{\sigma\nu}^\rho
   - \Gamma^\sigma_{\rho\nu} \tilde{J}_{\mu\sigma}^\rho &=& \nonumber \\
   && \hspace{-12em} - \, \frac{8\pi G}{c^4} \Big( {R^\sigma}_{\rho\mu\nu} {T_\sigma}^\rho
   + R_{\mu\alpha} {T_\nu}^\alpha - {T_\mu}^\alpha R_{\alpha\nu} \Big) , \qquad
\label{mainresult}
\end{eqnarray}
where the $R$ tensors characterize the curvature associated with the metric $g_{\mu\nu}$. The term involving the Riemann curvature tensor ${R^\sigma}_{\rho\mu\nu}$ vanishes for symmetry reasons: if the index $\sigma$ on the Riemann tensor is lowered by the metric and raised by the inverse metric on the energy-momentum tensor, the resulting tensors are antisymmetric and symmetric in $\rho$ and $\sigma$, respectively. The condition for the terms involving the Ricci tensor $R_{\mu\nu}={R^\sigma}_{\mu\sigma\nu}$ to cancel is given by the commutation relation
\begin{equation}\label{condcommu}
   R_{\mu\alpha} {T_\nu}^\alpha = {T_\mu}^\alpha R_{\alpha\nu} .
\end{equation}
This condition is satisfied whenever the Ricci tensor is a local function of the energy-momentum tensor. For example, Einstein's field equation constitutes a linear relation between Ricci tensor and energy-momentum tensor. More generally, the commutation relation (\ref{condcommu}) may be regarded as a much weaker ``no-conflict condition.'' If this condition is fulfilled, the currents (\ref{Jtildedef}) satisfy the conservation laws (\ref{Jtildeconservation}), which are equivalent to those in Eq.~(\ref{Jaconservation}).

For the purpose of illustration, we next consider the static isotropic solution around a mass point with mass $M$ at rest at the origin. In the symmetric gauge, we assume that the tetrad variables are of the form
\begin{equation}\label{isoxb}
   {b^\kappa}_\mu = \left( \begin{matrix}
   b & 0 \\
   0 & a \left( \delta_{km} - \frac{x_k x_m}{r^2} \right)  + q \frac{x_k x_m}{r^2}
   \end{matrix} \right) ,
\end{equation}
with suitable functions $a$, $b$ and $q$ of the distance from the origin, $r$. The field equations (\ref{YMfieldeqs}) for the gauge vector fields (\ref{compositionrule}) imply two evolution equations containing third derivatives of $a$ and $b$, respectively. By constructing the Robertson expansions of the solutions in terms of powers of $r_0/r$, where $r_0=GM/c^2$ is a characteristic length scale of the order of the Schwarzschild radius, one realizes that the field equations characterize the functions $a$ and $b$ with the only exception of the $1/r$ contribution to $a$. The function $q$ remains completely undetermined, except for the $1/r$ term. The Robertson expansions of the solutions for $a$ and $b$ imply that these functions can equivalently be described by remarkably simple first-order differential equations. This simplification is a consequence of the closed form expressions of the gauge-vector fields revealed in Eq.~(39) of \cite{hco231}.

The commutation relation (\ref{condcommu}) is satisfied automatically for the static isotropic solution. The component ${T_0}^0$ is the only non-vanishing component of the energy-momentum tensor. For the metric implied by Eq.~(\ref{isoxb}), the components $R_{0n}=R_{n0}$ of the Ricci tensor vanish. Therefore, no conflict can arise between the temporal component of the energy-momentum tensor and the spatial components of the Ricci tensor. This intuitive argument can be confirmed by a direct verification of Eq.~(\ref{condcommu}). We therefore need an additional equation to obtain a complete solution for the static isotropic field.

The additional equation plays the role of a coordinate condition. As the composite theory of gravity \cite{hco231}, just like the underlying Yang-Mills theory, is formulated in a background Minkowski space-time, there arises the problem of how to characterize the ``good'' coordinate systems in which the theory can be applied. This characterization should be Lorentz invariant, but not invariant under general coordinate transformations, that is, it shares the formal properties of coordinate conditions in general relativity. However, the unique solutions obtainable from Einstein's field equations only after specifying coordinate conditions are all physically equivalent, whereas the coordinate conditions in composite gravity actually characterize the physically preferred systems. From a historical perspective, it is remarkable that Einstein in 1914 still believed that the metric should be completely determined by the field equations and, therefore, a generally covariant theory of gravity was not desirable (see \cite{Giovanelli21} for a detailed discussion).

For our discussion of the static isotropic solution, we here propose the coordinate conditions
\begin{equation}\label{cocos}
   \frac{\partial^2}{\partial x^\sigma \partial x_\sigma}
   \left( \frac{\partial g_{\mu\rho}}{\partial x_\rho}
   - \frac{1}{2} \, \frac{\partial {g_\rho}^\rho}{\partial x^\mu} \right) = 0 ,
\end{equation}
for $r>0$. This is the linearized version of harmonic coordinate conditions raised to the level of third-order differential equations, just like the field equations. For practical purposes, this condition can be expressed as a first-order differential equation for $q$. In more general situations, more sophisticated coordinate conditions may turn out to be required. The guiding principles for formulating general coordinate conditions are given by (i) ensuring unique solutions, (ii) respecting the ``no-conflict condition'' (\ref{condcommu}), and (iii) supporting a Hamiltonian formulation of third-order field equations with proper auxiliary fields along the lines described in \cite{hco237,hco240}.

The Robertson expansions for the solutions of the field equations and coordinate conditions, which actually reproduce the high-precision predictions of general relativity, have one free parameter, namely the coefficient of the $(r_0/r)^3$ contribution to $q$. By proper choice of this coefficient (around $0.8$), the numerical solution obtained by integrating from large $r$, where the Robertson expansion is valid, to small $r$ can be kept free of any divergences for $r>0$. The required fine-tuning of the free parameter can be supported by shooting at the proper asymptotic solution for small $r$,
\begin{eqnarray}
   a(r) &=& a_0 \frac{r_0}{r} + \frac{2}{3} \sqrt{\frac{2r}{7r_0}} , \quad
   b(r) = -2 \sqrt{\frac{2r_0}{7r}} , \nonumber \\
   q(r) &=& \left( 1 + \frac{r}{r_0} \right)  \sqrt{\frac{2r_0}{7r}} .
\label{abqsingat0}
\end{eqnarray}

\begin{figure}
\centerline{\includegraphics[width=7.5 cm]{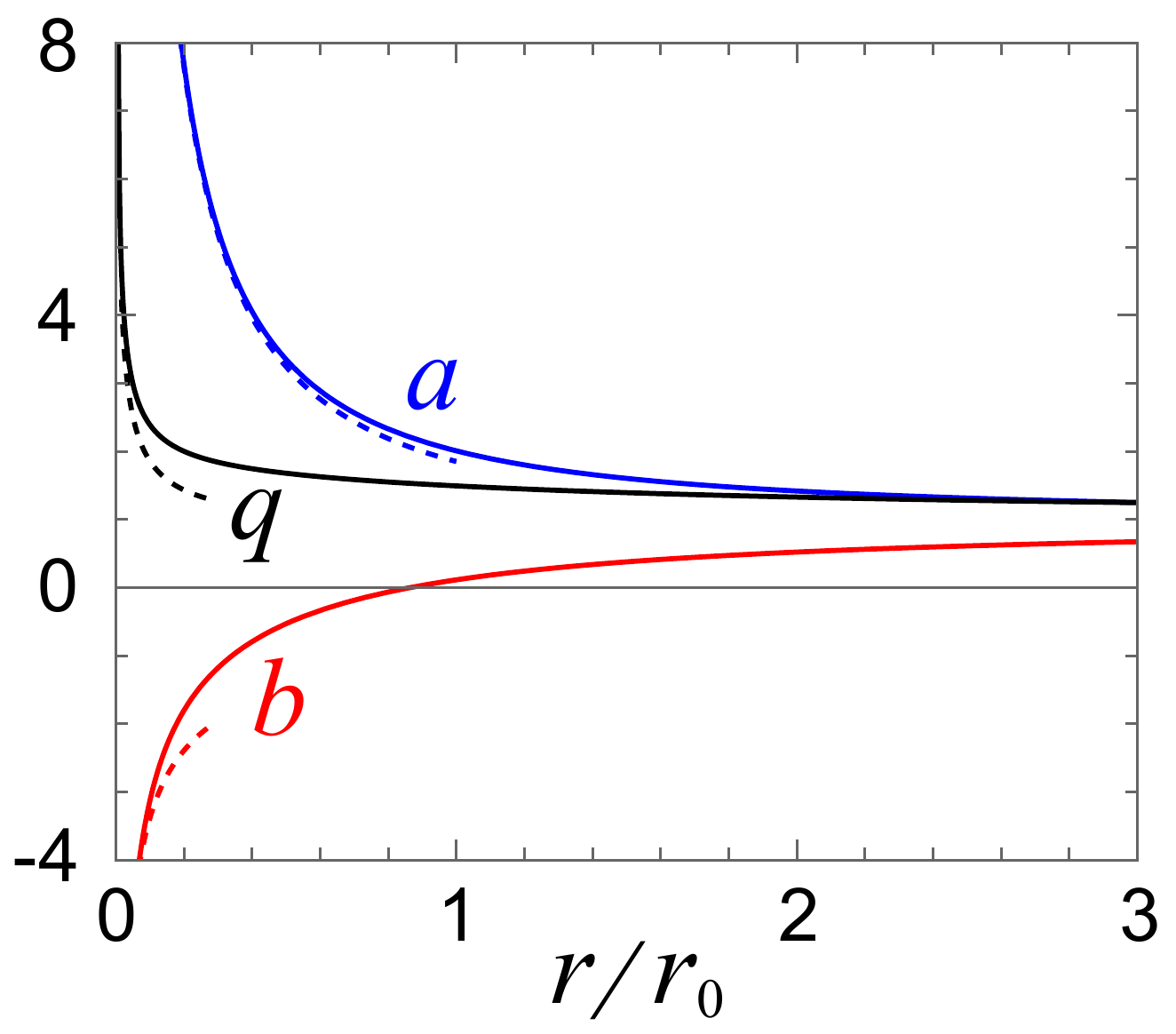}}
\caption[ ]{The functions $a$, $b$ and $q$ characterizing the tetrad variables (\ref{isoxb}) in the symmetric gauge for the static isotropic solution around a point mass at rest obtained from the composite theory of gravity. The asymptotic behavior (\ref{abqsingat0}) at $r=0$ is indicated by the respective dashed curves, where $a_0=1.496$ is the only fit parameter.} \label{YM_currents_fig_isotropic}
\end{figure}

The numerical solution for the static isotropic field is shown in Figure~\ref{YM_currents_fig_isotropic}. A noteworthy feature of this solution is that $b$ changes sign near the Schwarzschild radius. For the metric, this simply means that $g_{00} = - b^2$ touches the value zero near $r_0$. At that distance from the origin, proper time does not change with background time. The lack of singularities at finite $r$ is crucial when working in a background Minkowski space because, unlike in general relativity, one cannot remove singularities by general coordinate transformations.

In conclusion, we have provided the conserved currents (\ref{Jtildedef}) for the Yang-Mills-type gauge-field theory with Lorentz symmetry group in terms of the energy-momentum tensor. It is essential that we express the gauge vector fields in terms of tetrad variables, such that the local gauge symmetry arises from the freedom in decomposing the metric field characterizing an anisotropic velocity-momentum relation. The conserved currents provide the coupling of the gauge vector fields to matter. If the gauge vector fields are expressed in terms of the tetrad variables, we obtain the composite theory of gravity \cite{hco231,hco240}, which is a higher derivative theory (third-order differential equations for the tetrad variables). Constraints play an important role for avoiding instabilities, possibly also for eliminating potential problems associated with the non-compact nature of the Lorentz group, and for reducing the too large number of degrees of freedom. The commutation relation (\ref{condcommu}) guaranteeing the existence of conserved currents may be regarded as an extra constraint, if not already implied by the existing ones resulting from the composition rule (\ref{compositionrule}) or from the gauge conditions.

There exists a canonical Hamiltonian formulation of the evolution equations of our higher derivative, composite Yang-Mills theory in the combined space of tetrad and gauge-vector fields as configurational variables and their conjugate momenta \cite{hco237}. This is a clear advantage for the quantization of the theory. The constraints resulting from the composition rule are found to be gauge invariant, second class constraints. This suggests that, in the quantization process, they can be treated via Dirac brackets \cite{Dirac50,Dirac58a,Dirac58b}, and the gauge constraints can be treated independently with the usual BRST procedure \cite{BecchiRouetStora76,Tyutin75,Nemeschanskyetal86}. Therefore, if the composite Yang-Mills-type theory with Lorentz symmetry group should be recognized as a valid theory of gravity, its quantization in the context of dissipative quantum field theory \cite{hcoqft} appears to be straightforward.

%\bibliography{hcopubs}

\end{document}